\documentclass[%
reprint,
superscriptaddress,
 amsmath,amssymb,
 aps,
prl
]{revtex4-1}

\usepackage{graphicx}
\usepackage{dcolumn}
\usepackage{bm}

\usepackage[usenames]{xcolor}

\begin{document}

\title{Gravitational search for near Earth black holes or other compact dark objects}
\author{Tomoyo Namigata}
\email{tnamigat@iu.edu}
\affiliation{Center for Exploration of Energy and Matter and Department of Physics, Indiana University, Bloomington, IN 47405, USA}

\author{C. J. Horowitz}
\email{horowit@indiana.edu}
\affiliation{Center for Exploration of Energy and Matter and Department of Physics, Indiana University, Bloomington, IN 47405, USA}

\author{R. Widmer-Schnidrig}
\email{widmer@gis.uni-stuttgart.de}
\affiliation{Black Forest Observatory (BFO), Heubach 206, D-77709 Wolfach, Germany} \affiliation{Institute of Geodesy, University of Stuttgart, Stuttgart, Germany}

\date{\today}

\begin{abstract}
Primordial black holes, with masses comparable to asteroids, are an attractive possibility for dark matter.  In addition, other forms of dark matter could form compact dark objects (CDO).  We search for small tidal accelerations from low mass black holes or CDOs orbiting near the Earth, and find none.  Using about 10 years of data from the superconducting gravimeters in the Black Forest Observatory in South-Western Germany and at Djougou, Northern Benin in Western Africa we set an upper limit on the maximum mass of any dark object orbiting the Earth as a function of orbital radius.  For semi-major axis less than two earth radii we exclude all black holes or CDOs with masses larger than $6.7\times 10^{13}$ kg.  Lower mass primordial black holes may be strongly constrained by Hawking radiation.  We conclude that near Earth black holes are extremely unlikely.

\end{abstract}

\maketitle
Primordial black holes (PBH), made by density fluctuations in the early universe, are an attractive candidate for dark matter \cite{Green_2021,doi:10.1146/annurev-nucl-050520-125911}.  These PBH could have a large range of possible masses depending on the scale of the density fluctuations.  However, dark matter made of PBH with masses $M_{PBH}<10^{14}$ kg is constrained by Hawking radiation \cite{PhysRevLett.126.171101} while $M_{PBH}>10^{19}$ kg is constrained by microlensing \cite{microlensing2019,Alcock:1995dm,Alcock:2000ph,Tisserand:2006zx,Paczynski:1986}.  This leaves a very interesting open window of PBH masses between $10^{14}$ and $10^{19}$ kg that is largely unconstrained by present observations.  For example, De Luca et al. explain NANOgrav observations \cite{Arzoumanian:2020vkk} with gravitational waves from the formation of PBH with masses in this range \cite{PhysRevLett.126.041303}.  

There could be black holes in the solar system.   Indeed, Scholtz and Unwin speculate that Planet 9, a possible outer solar system object of several Earth masses, is a PBH \cite{PhysRevLett.125.051103,Siraj_2020}.  If dark matter is composed of PBH, we expect $\approx 10^3 (10^{14}{\rm\, kg} /M_{PBH})$ PBH to be present in the solar system (of volume $\approx10^6$ AU$^3$) at any given time.  It is possible that one of these objects could approach Earth or even be captured into orbit through a three body interaction.  

More generally, it is possible that Earth may have one or more ``dark moon" in addition to Luna.  These objects could be composed of PBH or other forms of dark matter that allows them to escape detection via electromagnetic observations.   In addition to PBH, other forms of dark matter may concentrate into compact dark objects (CDO).  These objects are assumed to have small non-gravitational interactions with normal matter.  Some possibilities or names for CDO include Boson Stars \cite{Boson_stars}, Dark Blobs \cite{PhysRevD.98.115020}, asymmetric dark matter nuggets \cite{PhysRevD.97.036003}, Exotic Compact Objects \cite{Giudice_2016}, Ultra Compact Mini Halos (UCMH) \cite{PhysRevD.85.125027} made for example of axions \cite{Yang:2017cjm}, and Macros \cite{10.1093/mnras/stv774}.      

The lifetime of near Earth orbits for conventional objects is short because of air resistance.   Dark objects, be they PBH or CDO, may have unique long lived near Earth orbits because of their small non-gravitational interactions. 
In this letter, we directly search for small gravitational accelerations (tides) from a PBH or CDO, with a $10^{14}-10^{19}$ kg mass, in a near Earth orbit.

In previous work we searched for gravitational waves from CDO merging with neutron stars \cite{Horowitz_Reddy} or orbiting inside the sun \cite{Horowitz:2019pru}.  We also searched for small gravitational accelerations from CDO moving inside the Earth \cite{PhysRevLett.124.051102}, see also \cite{Budker,Figueroa_2021,Tino_2021}.  In the present letter we significantly extend ref. \cite{PhysRevLett.124.051102} to search for CDO or PBH orbiting partially or completely outside the Earth with a range of longer orbital periods.


One might observe electromagnetic radiation as material accretes onto a black hole, see for example \cite{Siraj_2020}. Even if there is little accretion, one can still search for a dark object's gravity by accurately tracking spacecraft, see for example \cite{witten2020searching,lawrence2020bruteforce}, or using a gravimeter on Earth.

Gravimeters \cite{doi:10.1063/1.1150092} measure the local acceleration due to gravity and can detect small tides from an orbiting CDO or PBH.  Sensitive superconducting gravimeters have been deployed at several locations around the world \cite{IGETS}. They are used to observe a wide range of geophysical phenomena including Chandler wobble, solid Earth tides, post glacial rebound, seismic free oscillations and hydrological processes \citep{HindererCrossleyWarburton:07}.
In addition to geophysics, they have been used to search for a dependence of gravity on a hypothetical preferred reference frame \cite{1972ApJ...177..757W,1976ApJ...208..881W}, or the violation of Lorentz invariance \cite{PhysRevLett.119.201101,PhysRevD.97.024019}, as the Earth translates or rotates.  In addition, gravimeters have been used to search for oscillations of the Earth excited by gravitational waves \cite{PhysRevD.90.042005}.

An object orbiting through or around the Earth with coordinate ${\bf r}(t)$ will have an acceleration,
\begin{equation}
\frac{d^2{\bf r}(t)}{dt^2}=-G\frac{M_{enc}(r)}{r^3}{\bf r}\, .
\label{eq.racc}
\end{equation}
Here $G$ is Newton's constant and $M_{enc}(r)$ is the enclosed mass of the Earth that is interior to $r$.  This reduces to the mass of the Earth $M_E$ for $r>R_E$ the radius of the Earth, see ref. \cite{PhysRevLett.124.051102} for details.  We numerically integrate Eq. \ref{eq.racc} using the Velocity Verlet algorithm with a time step $\Delta t=1$ s, typically for a total time of $2^{26}$ s.  This simple procedure works for orbits inside, partially inside, and outside the Earth.  We assume the Earth is spherically symmetric.  Explicitly including the Earth's quadrupole moment does not significantly change results for excluded masses, see below.  We also neglect small perturbations from the moon or the sun.  These are expected to be unimportant for $r$ near the Earth.


A gravimeter located at ${\bf R}(t)$ will feel a time-dependent acceleration $a(t)$ that is the difference of the gravitational acceleration from the compact object minus the acceleration of Earth towards the object,  
\begin{equation}
a(t)=GM_D\Bigl\{\frac{M_{enc}(r)}{M_E\, r^3}{\bf r}-\frac{{\bf r}-{\bf R}}{|{\bf r}-{\bf R}|^3}\Bigr\}\cdot\hat {\bf R} .
\label{eq.a}
\end{equation}
Here $M_D$ is the mass of the dark object, either $M_{PBH}$ or $M_{CDO}$, and $\hat{\bf R}={\bf R}/R$.  Note that the gravimeter only measures the vertical component of ${\bf a}$.  We choose a coordinate system where the gravimeter is located at ${\bf R}(t)=R_E(\cos\theta_l\cos\omega t,\cos\theta_l\sin\omega t, \sin\theta_l)$.  Here $\theta_l$ is the latitude of the gravimeter (48.33$^\circ$ N for the Black Forest Observatory) and $\omega=2\pi/{\rm day}$.   We fast Fourier transform (FFT) $a(t)$ from Eq. \ref{eq.a} (including a Hanning taper) using the numerically integrated orbit ${\bf r}(t)$. 
This yields the signal amplitude $A(f)$ for frequency $f$ that will be compared to the FFT of gravimeter data.  We will be interested in $f$ from $\approx 10^{-6}$ to $2\times 10^{-4}$ Hz depending on the orbit.

We now analyze gravimeter data.  Data from super-conducting gravimeters (SGs), deployed at various locations around the world, has been archived by the Global Geodynamics Project (GGP, 1997-2015) and by the International Geodynamics and Earth Tide Service (IGETS, 2015-) \cite{IGETS,gravimeter_data}. We focus on the SG at the Black Forest Observatory (BFO at 48.33$^\circ$N, 8.33$^\circ$E) in South Western Germany because of its low noise.  We also consider the SG at Djougou, Northern Benin (DJ at 9.74$^\circ$N, 1.61$^\circ$E) in West Africa because of its location near the equator.  

We analyze 11.1 years of BFO gravimeter data from Dec. 1, 2009 to Jan. 23, 2021 \cite{BFO_data} and 7.9 years of DJ data from Jan. 1, 2011 to Nov. 30, 2018 \cite{DJ_data}.  We analyze this data as in Ref. \cite{PhysRevLett.124.051102}, but now focusing on the lower frequency band $\approx 10^{-6}$ to $2\times 10^{-4}$ Hz.  We carefully handle artifacts in the data: times when the instrument behaved non-linearly due to saturation from large quakes, operator interference or other malfunctions.  We subtract from the data a synthetic tidal model for the stations that includes the effect of ocean loading \citep{ETERNA}.  In addition we partially correct for accelerations due to atmospheric mass fluctuations above the gravimeter based on the barometric pressure recorded near the gravimeter, see Ref. \cite{PhysRevLett.124.051102} for details.

\begin{figure}[htb]
\includegraphics[width=1\columnwidth]{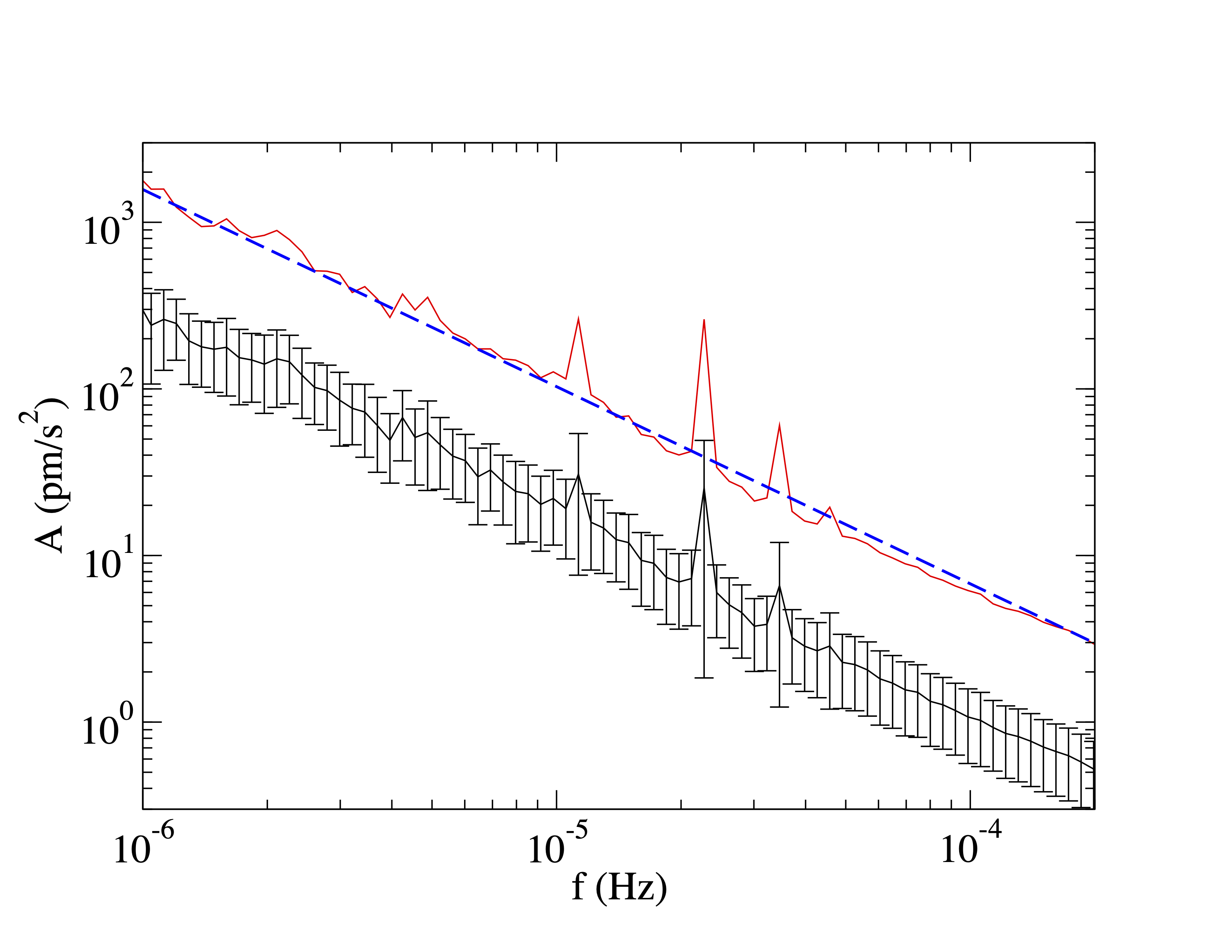}
 \caption{ 
Fourier amplitude spectra of gravity residuals for the Black Forest Observatory gravimeter versus frequency $f$.  Shown in black are the mean and standard deviation $\sigma$.  The red line is a conservative upper limit of mean plus 10$\sigma$ and the dashed blue line is the simple fit to this limit given by Eq. \ref{Eq.BFOlimit}.  The three peaks in the upper limit at frequencies above $10^{-5}$ Hz are solar and lunar diurnal, semidiurnal, and terdiurnal residual tides.}
\label{Fig1}
\end{figure}

\begin{figure}[htb]
\includegraphics[width=1\columnwidth]{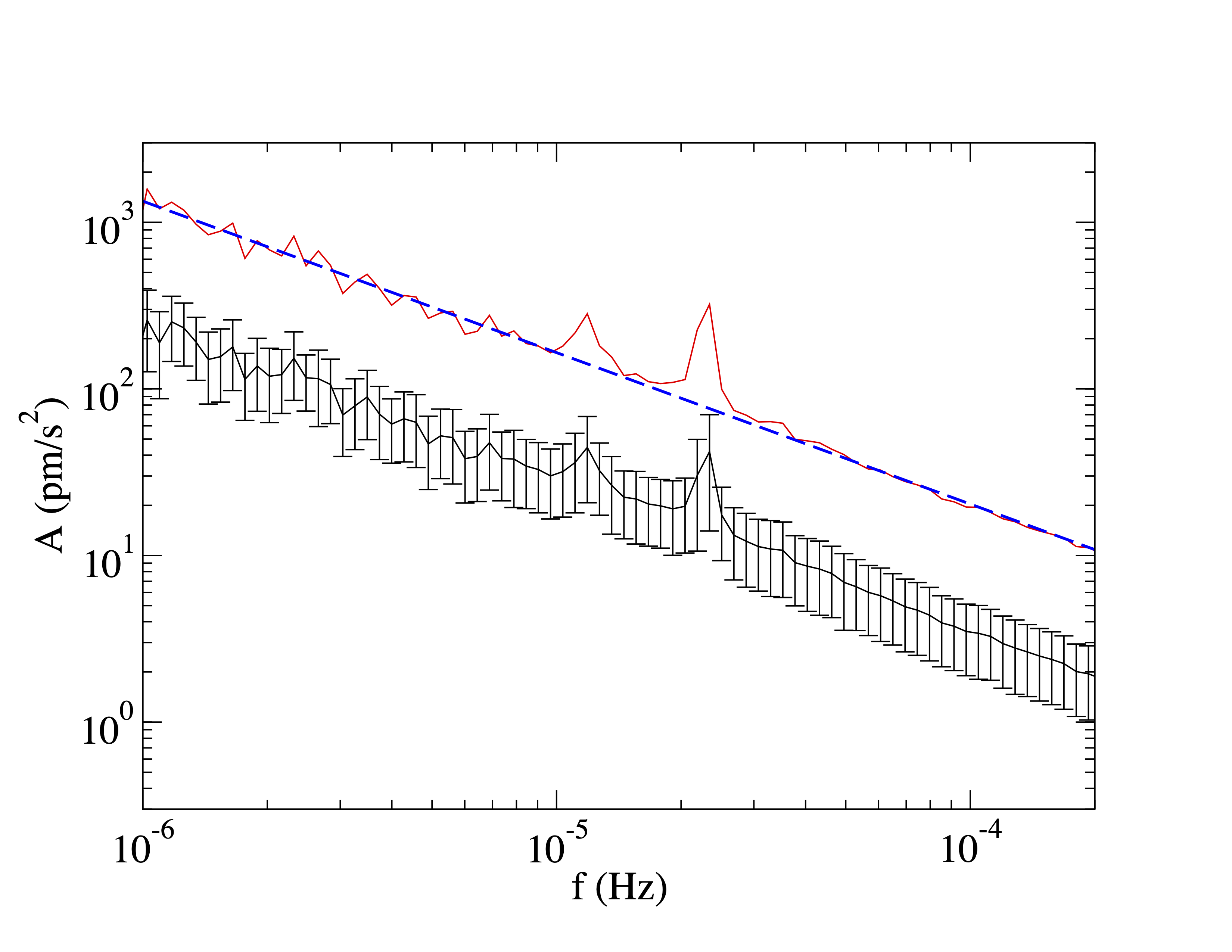}
 \caption{ 
Fourier amplitude spectra for the Djougou gravimeter as in Fig. \ref{Fig1}.  The dashed blue line is the fit to the upper limit in Eq. \ref{Eq.DJlimit}.
\label{Fig1b}}
\end{figure}

In Figs. \ref{Fig1} and \ref{Fig1b} we show Fourier amplitude spectra of the pressure corrected and Hanning tapered gravity residuals for BFO and DJ, respectively.  These spectra are normalized such that a pure time domain signal $h(t)=A_0\sin(2\pi f_0t)$ of amplitude $A_0$ yields $A(f_0)=A_0$.  The average noise and standard deviation for a number of frequency bins is shown along with a conservative upper limit that is 10$\sigma$ above the average noise.  A signal from a PBH above this limit can confidently be ruled out \footnote{If a dark object signal happened to have exactly the same frequency as one of the lunar or solar tidal lines then the object could ``hide'' in the large tidal background.  This provides a small exception.}.  We fit these upper limit noise curves with
\begin{equation}
    A_{\rm BFO}(f) \approx 1.25\times 10^{-4} \Bigl(\frac{1\ \rm Hz}{f}\Bigr)^{1.1833}\ {\rm pm/s^2}
    \label{Eq.BFOlimit}
\end{equation}
for the BFO station and 
\begin{equation}
    A_{\rm DJ}(f) \approx 4.71\times 10^{-3} \Bigl(\frac{1\ \rm Hz}{f}\Bigr)^{0.909} \ {\rm pm/s^2}
    \label{Eq.DJlimit}
\end{equation}
for the DJ station.  These fits are valid for $10^{-6}$ Hz $ <f<2\times 10^{-4}$ Hz.

\begin{figure}[htb]
\includegraphics[width=\columnwidth]{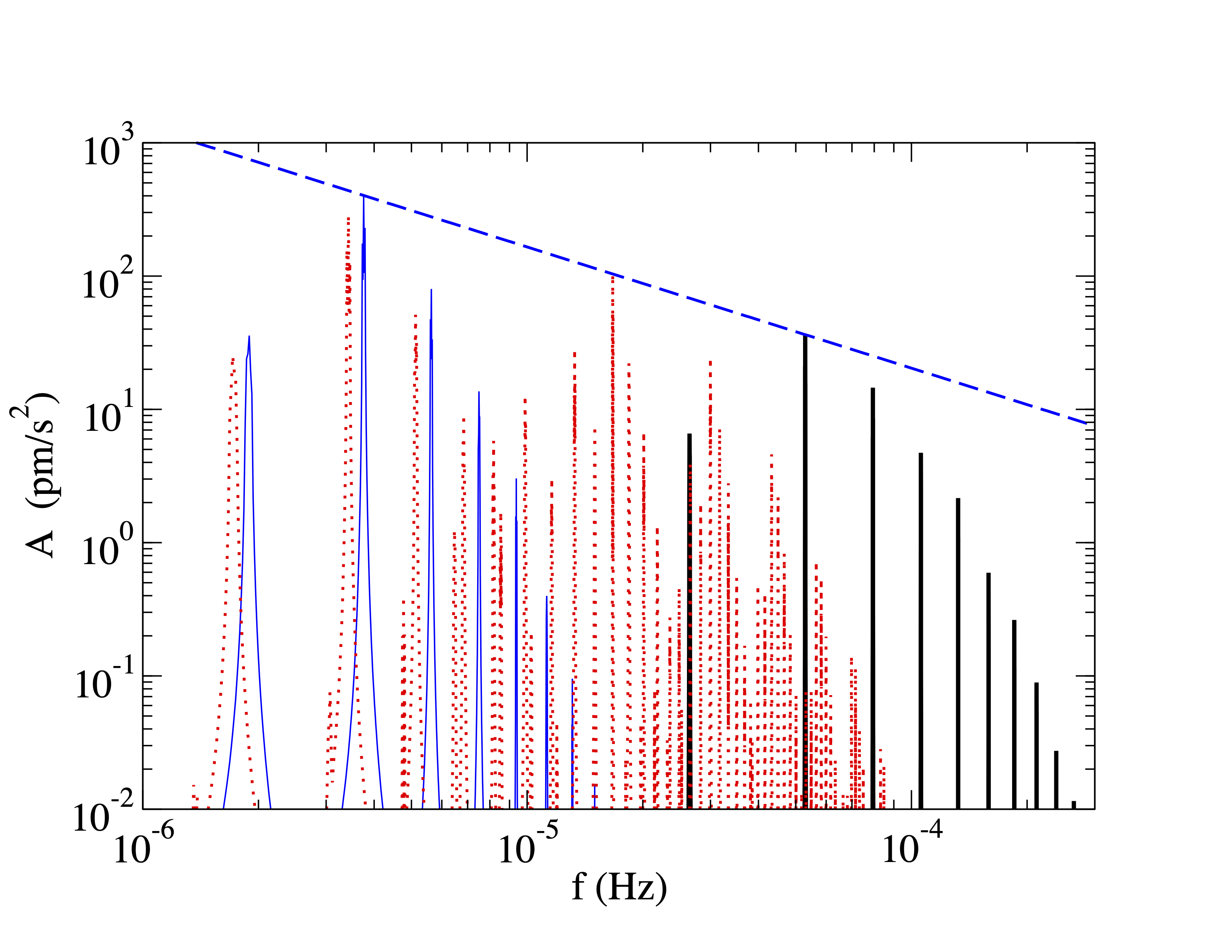}
 \caption{Fourier amplitude $A(f)$ vs frequency $f$ for equatorial (inclination $\theta_I=0$) orbits observed with the DJ gravimeter. Eccentricity $\epsilon=0$ orbits with semimajor axis $a=3R_E$ (thick black curve) and $a=6R_E$ (blue curve) are shown in addition to an elliptical orbit with $a=6.05R_E$ and eccentricity $\epsilon=0.1$ (dotted red curve). In each case the object mass is equal to the appropriate excluded mass see Table \ref{Table1}.  The dashed blue curve shows the DJ noise limit from Eq. \ref{Eq.DJlimit}.}
\label{Fig2}
\end{figure}

\begin{figure}[htb]
\includegraphics[width=1.0\columnwidth]{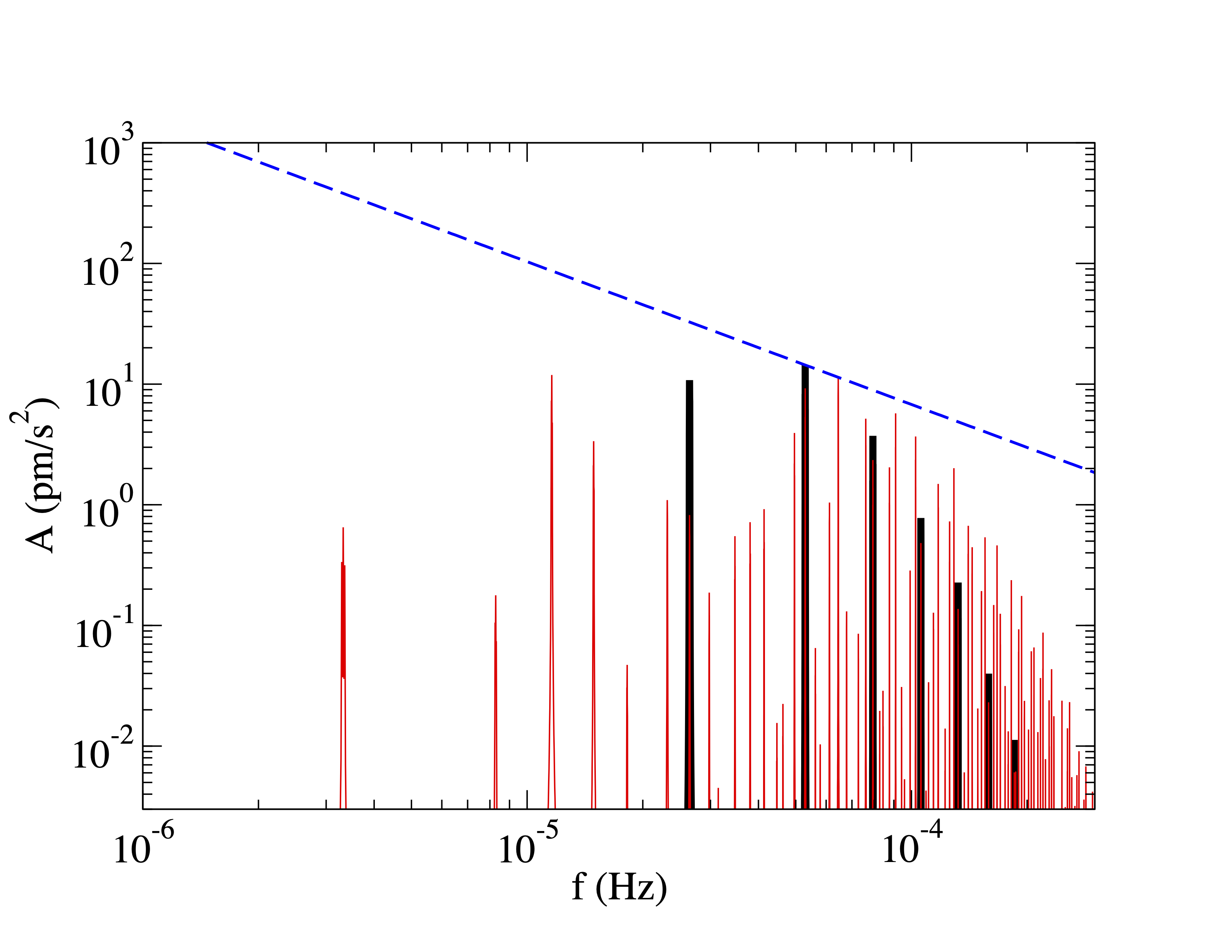}
 \caption{ 
Fourier amplitude $A(f)$ vs frequency $f$ for circular orbits of radius $a=3R_E$ as observed by the BFO gravimeter. The thick black lines are for inclination $\theta_I=0^\circ$ and the thin red lines are for $\theta_I=45^\circ$.  In each case the object mass is equal to the appropriate excluded mass see Table \ref{Table1}.  The blue dashed curve shows the BFO noise limit from Eq. \ref{Eq.BFOlimit}. }
\label{Fig3}
\end{figure}

We compare Fourier amplitude spectra $A(f)$ for a given orbit to the gravimeter upper limit noise curves in Eqs. \ref{Eq.BFOlimit}, \ref{Eq.DJlimit} to deduce excluded masses $M_{ex}$.  The excluded mass is the smallest object mass $M_D$ in Eq. \ref{eq.a} so that $A(f)$ just equals the noise limit curve for one harmonic frequency $f$.  Any object with $M_D>M_{ex}$ can be ruled out by the gravimeter data.  Table \ref{Table1} presents excluded masses for some selected orbits as shown in Figs. \ref{Fig2},\ref{Fig3}.  The final excluded mass for a given orbit is the minimum of the excluded mass based on BFO and DJ observations.

\begin{table}[hbt]
\begin{tabular}{ccccc}
$a$ ($R_E$) &  $\theta_I$ ($^\circ$)   & $\epsilon$ & Gravimeter & $M_{ex}$ (kg) \\
\hline
\rule{0pt}{2.8ex}3 & 0 & 0 & DJ & $4.0\times 10^{14}$ \\
6 & 0 & 0 & DJ & $3.7\times 10^{16}$ \\
6.05 & 0 & 0.1 & DJ & $2.7\times 10^{16}$ \\
3 & 0 & 0 & BFO & $5.0\times 10^{14}$\\
3 & 45 & 0 & BFO & $3.0\times 10^{14}$ \\
\end{tabular}
\caption{\label{Table1} Excluded mass $M_{ex}$ for example orbits with semimajor axis $a$, inclination $\theta_I$, and eccentricity $\epsilon$ as observed with the indicated gravimeter.}
\end{table}

Figure \ref{Fig2} shows $A(f)$ for DJ observations of equatorial orbits (inclination $\theta_I=0$).  As the semimajor axis $a$ of the orbit increases, $A(f)$ shifts to lower frequencies, where the gravimeter noise is larger. This shift is because of the larger orbital period.  Elliptical orbits (with nonzero eccentricity $\epsilon$) have higher harmonics with larger amplitudes compared to $A(f)$ for a circular orbit.  As a result $M_{ex}$ for $\epsilon>0$ is expected to be lower than $M_{ex}$ for a circular orbit (with the same $a$).  This can be seen in another way. Elliptical orbits come closer to the gravimeter than do circular orbits, with the same $a$.  This increases the gravity signal of the compact object and leads to a smaller $M_{ex}$.   

Figure \ref{Fig3} shows $A(f)$ for BFO observations of circular orbits.  As the inclination of the orbit $\theta_I$ increases from 0 to 45$^\circ$, $M_{ex}$ decreases.  In general, inclined orbits will come closer to the BFO gravimeter, at latitude 48.33$^\circ$ N, than do equatorial orbits. This leads to larger $A(f)$, for a given $M_D$, and a smaller excluded mass $M_{ex}$. Furthermore, inclined orbits will have more harmonics in $A(f)$ involving  sums and differences of the orbital frequency and multiples of the Earth's rotational frequency.  This is also illustrated in Fig. \ref{Fig3}. 


\begin{figure}[htb]
\includegraphics[width=1.0\columnwidth]{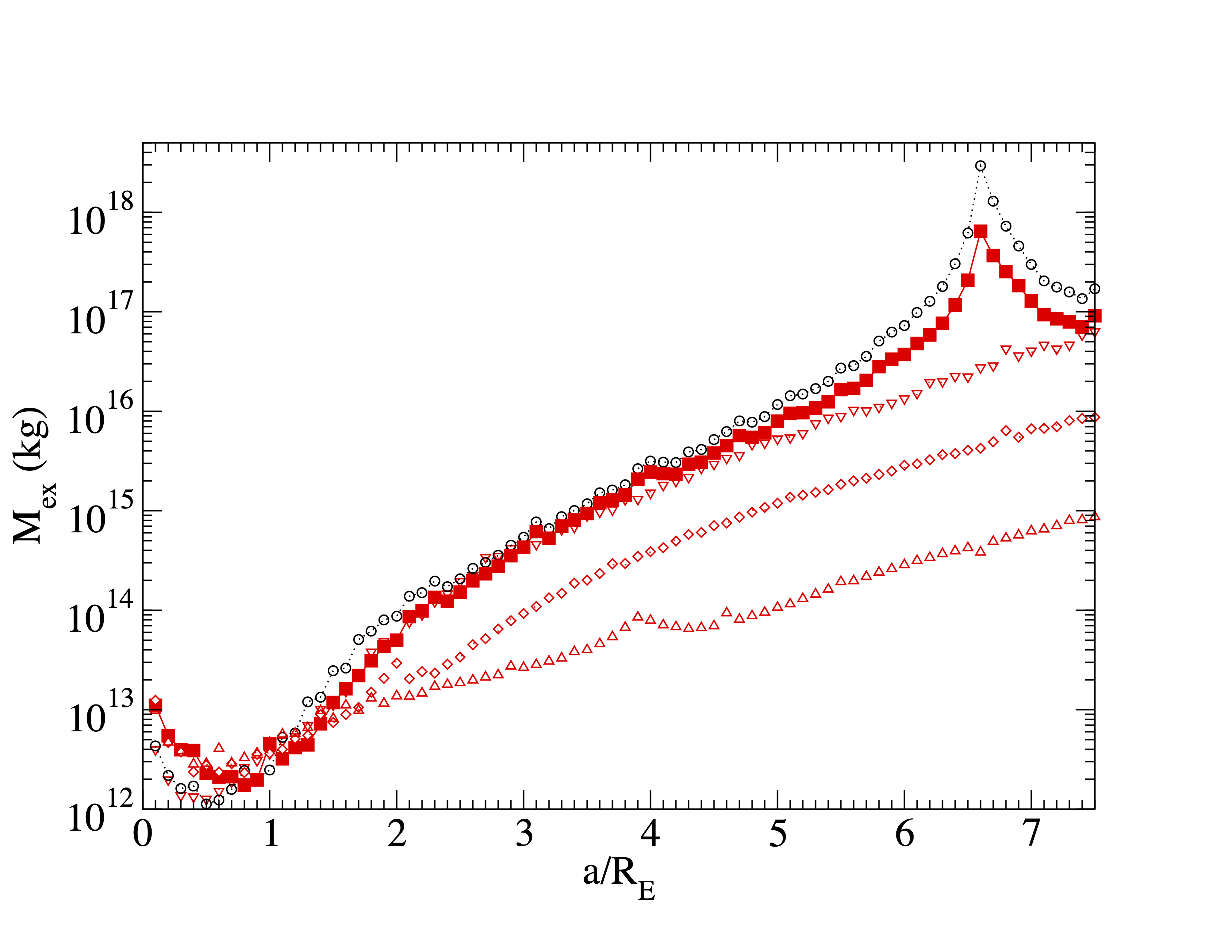}
 \caption{ 
Excluded mass $M_{ex}$ versus $a$ for equatorial orbits $\theta_I=0$.  Black circles are for BFO, and red squares DJ, observations of circular orbits.  Elliptical orbits observed with DJ are shown for eccentricities $\epsilon=0.25$ downward red triangles, $0.5$ diamonds, and $0.75$ upward triangles.}
\label{Fig4A}
\end{figure}

\begin{figure}[htb]
\includegraphics[width=1.0\columnwidth]{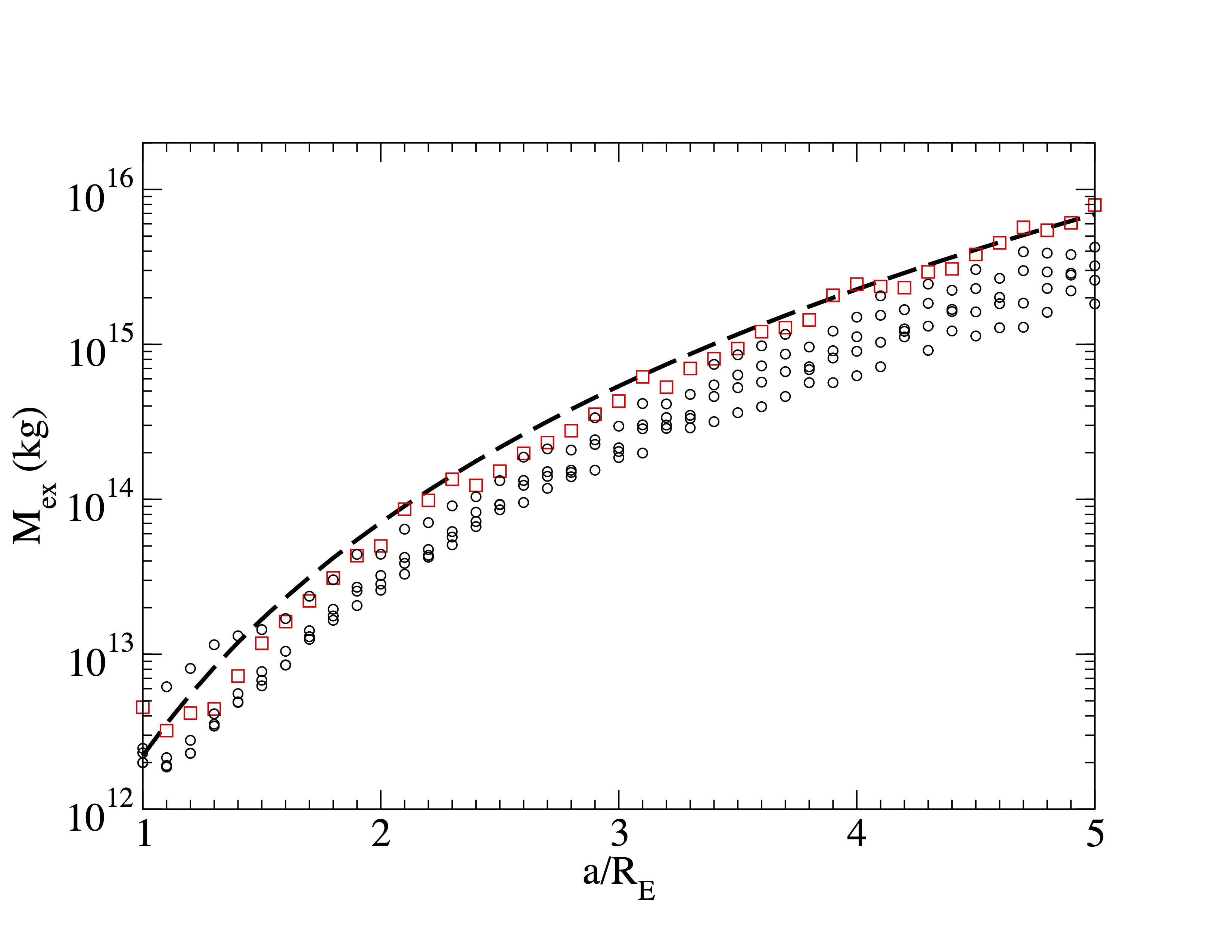}
 \caption{ 
Excluded mass $M_{ex}$ versus $a$ for circular orbits.  The red squares show DJ observations for $\theta_I=0$ while the black circles show BFO observations of inclined orbits with $\theta_I\geq 30^\circ$.   The dashed black line is $M_{ex}\approx 2.1\times 10^{12} (a/R_E)^5$ kg, see Eq. \ref{Eq.mex}.
\label{Fig4}}
\end{figure}

Figure \ref{Fig4A} shows excluded mass $M_{ex}$ versus $a$ for equatorial orbits $\theta_I=0$.  Here the DJ gravimeter dataset provides tighter constraints than the dataset from BFO.  Furthermore both gravimeters are much more sensitive for elliptical orbits than for circular orbits.

In general $M_{ex}$ increases smoothly with $a$.  However, there are resonances where the orbital period over the Earth's rotational period is equal to the ratio of small integers.  Resonances can introduce structure in $M_{ex}$ versus $a$.
There is a large peak in $M_{ex}$ for equatorial orbits with $a\approx 7R_E$.  This corresponds to nearly geostationary orbits. Here objects move very slowly with respect to the gravimeter.  As a result the gravimeter is almost blind to their presence.  This blind spot is only present for nearly equatorial circular orbits and largely vanishes by $\theta_I=15^\circ$. 

Excluded masses for $a<R_E$ in Fig. \ref{Fig4A} correspond to objects moving {\it inside} the Earth as discussed in ref. \cite{PhysRevLett.124.051102}. For these objects there is a second blind spot at the center of the Earth $a=0$.  Here the object is also at rest with respect to the gravimeter.

Figure \ref{Fig4} shows $M_{ex}$ for circular orbits with a range of inclinations. An approximate fit to an upper limit for all of the $M_{ex}$ in Fig. \ref{Fig4} is
\begin{equation}
M_{ex} \approx 2.1\times 10^{12} \bigl(\frac{a}{R_E}\bigr)^5 {\ \rm kg}\, .
\label{Eq.mex}
\end{equation}
This is shown as a dashed line in Fig. \ref{Fig4}.  Furthermore since $M_{ex}$ is smaller for elliptical orbits, Eq. \ref{Eq.mex} also serves as a conservative upper limit for elliptical orbits.  Therefore Eq. \ref{Eq.mex} provides an upper limit to $M_{ex}$ that is valid for orbits of any eccentricity or inclination.  It is our primary result. 
For example, at $a=2R_E$ we have $M_{ex}=6.7\times 10^{13}$ kg.  A spherical object of this mass and terrestrial densities has a radius of a few kilometers while a black hole of this mass has a Schwarzschild radius of 100 fm.

The $a^5$ dependence of $M_{ex}$ in Eq. \ref{Eq.mex} can be understood as follows.  First, the tidal force falls as $\approx 1/a^3$.  Second, the orbital frequency $f$ decreases as $1/a^{3/2}$ (Kepler's third law).  Finally, the noise limit in Eqs. \ref{Eq.BFOlimit} and \ref{Eq.DJlimit} increases as $f$ decreases.  Therefore the ratio of signal to noise decreases faster than  $1/a^3$ so that the excluded mass grows faster than $a^3$.

   
Our search is limited by the background in the gravimeter signal from atmospheric fluctuations.  It should be possible to somewhat increase the sensitivity with a coherent search involving gravimeters at several locations.  However, this would likely involve a great increase in complexity because one would need to analyze gravimeter data separately for every compact object orbit.

Searches around other solar system bodies are also possible.  The seismometer on the InSight lander on Mars has sensitivity as a gravimeter.  Unfortunately, this instrument is exposed to some wind and thermal noise \cite{https://doi.org/10.1029/2021EA001669}.  Observations on the Moon could largely avoid atmospheric fluctuations.  Indeed a gravimeter was deployed during Apollo 17 \cite{LunarSG,doi:10.1002/2014JE004724} but it did not function well. 

We searched for periodic gravimeter signals from small tidal interactions of dark objects and found none.  We rule out all near Earth objects orbiting with semimajor axis $a$ and mass $M >M_{ex}(a)$, see Eq. \ref{Eq.mex}.
We find $M_{ex}(2R_E)\approx 6.7\times 10^{13}$ kg and rule out all near Earth black holes orbiting with $a\leq2R_E$ and $M>6.7\times10^{13}$ kg.  Furthermore, dark matter made of primordial black holes of mass $M_{PBH}<10^{14}$ kg is significantly constrained by the lack of observation of Hawking radiation \cite{PhysRevLett.126.171101}.  Therefore a near Earth black hole may be extremely unlikely.

\begin{acknowledgments} 
We acknowledge helpful discussions with Dmitry Budker, Matt Caplan, Rafael Lang, Rees McNally, Cole Miller, Maria Alessandra Papa, Tanya Zelevinsky and Walter Z\"urn. CJH is supported in part by US Department of Energy grants DE-FG02-87ER40365 and DE-SC0018083. We gratefully acknowledge the work of the operators of the superconducting gravimeters: Jacques Hinderer and Fr\'ed\'eric Little from EOST, Strasbourg for the Djougou station and Thomas Forbriger and Peter Duffner from the Karlsruhe Institute of Technology for the Black Forest Observatory.  We thank the data centers for archiving and freely distributing the gravimetric data and Thomas Forbriger for porting the ETERNA software package for tidal predictions to UNIX \citep{ETERNA_UNIX} and Jean-Paul Boy from EOST Starsbourg for publishing his tidal analyses results.

\end{acknowledgments}
\bibliographystyle{apsrev}
\bibliography{DarkRattles}

\end{document}